# Circular currents in a magnetic ring with zero net magnetization in presence of a side-coupled one-dimensional chain

Sourav Karmakar[1[0009-0001-7027-0415]], Suparna Sarkar [2[0000-0003-4928-924X]], and Santanu K. Maiti[1[0000-0003-3979-8606]]

[1]*Physics and Applied Mathematics Unit, Indian Statistical Institute, 203 Barrackpore Trunk Road, Kolkata-700 108, India*
[2]*Theoretical Sciences Unit, Jawaharlal Nehru Centre for Advanced Scientific Research, Bangalore-560 064, India*
karmakarsourav2015@gmail.com, santanu.maiti@isical.ac.in

We investigate persistent charge and spin currents in a magnetic quantum ring threaded by an Aharonov-Bohm flux, in the presence of a side-coupled one-dimensional non-magnetic chain. The neighboring magnetic moments in the ring are arranged in an antiparallel configuration. In the absence of the chain, the spin circular current vanishes exactly due to the symmetry between the up and down spin sub-Hamiltonians. Modeling the system within a tight-binding framework, we compute the currents using a second-quantized approach. Both charge and spin currents can be selectively tuned by adjusting the ring-chain coupling strength. Temperature plays a crucial role in modulating the currents, and interestingly, we find that they increase significantly with rising temperature--contrary to conventional expectations.

## 1 Introduction

Flux-driven circular currents in small conducting loops are well-known phenomena that have been extensively studied over the years. Büttiker, Imry, and Landauer first proposed [1] that when a nanoscale ring encloses a magnetic flux φ, commonly referred to as the Aharonov-Bohm (AB) flux [2, 3], it sustains a net circular charge current. Interestingly, this current does not vanish even when the flux is removed, a phenomenon known as flux-driven *persistent* charge current (CC). This effect was first demonstrated experimentally by Levy and co-workers [4] using an array of copper rings. Since then, significant theoretical and experimental efforts have been devoted to exploring its various aspects.

Similar to persistent charge current, a persistent spin current (SC) can also arise under suitable conditions when spin-dependent scattering mechanisms are present. Common spin-dependent interactions in condensed matter systems include spin-orbit (SO) coupling [5, 6], Zeeman splitting, and spin-moment interactions [7, 8]. The latter is typically observed in magnetic systems, with ferromagnetic ring geometries being the primary focus of previous studies. However, recent investigations suggest that magnetic systems possessing vanishing net magnetization can also sustain persistent SC if the symmetry between the spin-up and spin-down sub-Hamiltonians is broken. Achieving such symmetry breaking is of course challenging, and in the present work, we propose a method to accomplish it. Given the notable advantages of such magnetic systems over their ferromagnetic counterparts [9, 10], there is growing interest in utilizing them for spin-dependent transport studies.

Here, we consider a tight-binding magnetic quantum ring subjected to a magnetic flux $\phi$ (measured in units of the elementary flux-quantum $\phi_0$), which breaks time-reversal symmetry and induces a charge current in the ring. To generate a spin current, we couple the ring to a non-magnetic (NM) one-dimensional (1D) chain (see Fig. 1). Both charge and spin currents are computed using a second-quantized formalism, where energy eigenstates rather than eigenvalues play a central role. The currents are analyzed at both zero and finite temperatures for a fixed chemical potential. Interestingly, at finite temperatures, an anomalous enhancement of both charge and spin currents is observed [11, 12], contrary to conventional expectations. This enhancement is directly linked to the ring-chain coupling and can be further modulated by tuning other physical parameters of the system. We discuss these effects in detail.

The rest of the paper is organized as follows. Section 2 describes the quantum system and the theoretical framework used for calculations. Section 3 presents and critically analyzes the numerical results. Finally, we conclude the key findings in Sec. 4.

## 2 Ring-wire coupled system and theoretical framework

### 2.1 Model quantum system and TB Hamiltonian

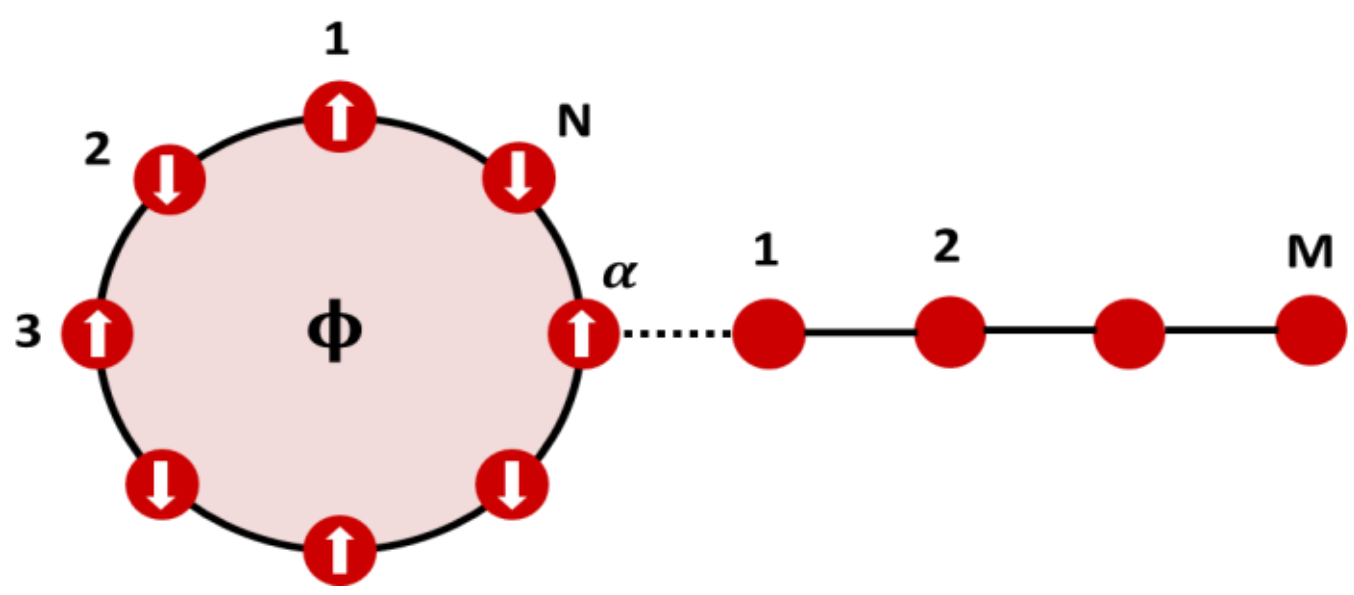

**Fig. 1.** Schematic diagram of a magnetic quantum ring with *N* sites coupled to a non-magnetic chain containing *M* sites. Each lattice site is represented by a filled colored ball. In the ring, the neighboring magnetic moments are oriented along ±z axes (spin quantized directions). The ring encloses a magnetic flux, which is responsible for generating the currents.

Let us begin with the ring-wire coupled system, schematically shown in Fig. 1, where a quantum ring is directly coupled to a non-magnetic chain. The neighboring magnetic moments are arranged in an anti-parallel configuration. We choose the total number of magnetic sites, *N*, in the ring to be even so that the net magnetization vanishes. The site $\alpha$ (variable) of the ring is directly coupled to site number 1 of the non-magnetic chain, possessing *M* number of lattice sites, via the coupling strength $\lambda$. A tight-binding (TB) framework is used to describe the system. Within the nearest-neighbor hopping (NNH) approximation, the TB Hamiltonian of the ring-wire coupled system reads as [13-17]

$$H = H_{Ring} + H_{Coupling} + H_{Chain}$$

$$= \sum_i [\boldsymbol{c}_i^\dagger (\boldsymbol{\epsilon}_i^R - \overrightarrow{\boldsymbol{h}_\iota}.\vec{\boldsymbol{\sigma}})\, \boldsymbol{c}_i + (\boldsymbol{c}_i^\dagger \boldsymbol{t}\, e^{j\theta} \boldsymbol{c}_{i+1} + h.c.)] + \left(\boldsymbol{c}_1^\dagger \boldsymbol{\lambda} \boldsymbol{c}_\alpha + h.c.\right)$$

$$+ \sum_i [\boldsymbol{c}_i^\dagger \boldsymbol{\epsilon}_i^C\, \boldsymbol{c}_i + (\boldsymbol{c}_i^\dagger \boldsymbol{t} \boldsymbol{c}_{i+1} + h.c.)] \qquad (1)$$

where the first, second, and last terms represent the sub-Hamiltonians of the ring, coupling between the ring and the chain, and the non-magnetic chain, respectively. Here $\boldsymbol{c}_i^\dagger = (c_{i\uparrow}^\dagger\ c_{i\downarrow}^\dagger)$ and $\boldsymbol{c}_i = \begin{pmatrix} c_{i\uparrow} \\ c_{i\downarrow} \end{pmatrix}$. $c_{i\sigma}^\dagger$ and $c_{i\sigma}$ are the conventional fermionic creation and annihilation operators, respectively, where $\sigma =\uparrow, \downarrow$. $\boldsymbol{\epsilon}_i^R$ =diag($\epsilon_{i\uparrow}\ \epsilon_{i\downarrow}$) and $\boldsymbol{t} =$ diag($t\ t$) are the site energy and hopping matrices, respectively. The phase factor $\theta$ is expressed as $\theta = 2\,\pi\phi/N\,\phi_0$. This phase appears due to the magnetic flux in the ring. The TB parameter $\epsilon_{i\sigma}$ denotes the site energy of an electron at the *i*-th site without any magnetic scattering effects, while $t$ represents the NNH strength. The term $\overrightarrow{\boldsymbol{h}_\iota}$ is a spin-dependent scattering factor, defined as the product of the spin-moment coupling strength J and the average local spin $< \overrightarrow{S_\iota} >$ i.e. $\overrightarrow{\boldsymbol{h}_\iota} = \mathrm{J} < \overrightarrow{\boldsymbol{S}_\iota} >$ [18]. Here, $\vec{\boldsymbol{\sigma}}$ represents the Pauli spin vector, with $\sigma_z$ being diagonal in our formulation. $\boldsymbol{\lambda} =$ diag($\lambda\ \lambda$), where λ measures the coupling strength between the ring and the chain. $\boldsymbol{\epsilon}_i^C$ =diag($\epsilon_i\ \epsilon_i$) are the site energies in the chain. We use the same parameter *t* to describe the electron hopping in the ring and the chain.

### 2.2 Calculation of charge and spin currents

In our study, we calculate both CC and SC using the second-quantization framework, which is the standard prescription, and it involves energy eigenstates instead of eigenvalues.

#### 2.2.1 Calculation of charge current

The charge current operator is defined as [19]

$$\mathbf{I}_c = \frac{e\dot{\mathbf{X}}}{Na} \qquad (2)$$

where $e$ represents the electronic charge, $\dot{\mathbf{X}}$ denotes the velocity operator, $N$ is the total number of lattice sites in the ring, and $a$ is the lattice spacing. The velocity operator can be represented using the position operator ($\mathbf{X}$) and the Hamiltonian as follows

$$\dot{\mathbf{X}} = \frac{1}{\mathrm{j}\hbar}[\mathbf{X}, \boldsymbol{\mathcal{H}}]. \tag{3}$$

The position operator can be written in terms of the fermionic operators as

$$\mathbf{X} = \sum_{m} \boldsymbol{c}_m^{\dagger} ma\, \boldsymbol{c}_m \tag{4}$$

where $m$ denotes the site index. Using the above relations, we obtain the charge current operator ($\mathbf{I}_{\mathrm{c}}$)

$$\mathbf{I}_{\mathrm{c}} = \frac{jet}{N\hbar}\sum_{i}(e^{-j\theta}\, c_{i\uparrow}^{\dagger}c_{i+1\uparrow} - e^{j\theta}c_{i+1\uparrow}^{\dagger}c_{i\uparrow}) + \frac{jet}{N\hbar}\sum_{i}(e^{-j\theta}\, c_{i\downarrow}^{\dagger}c_{i+1\downarrow} - e^{j\theta}c_{i+1\downarrow}^{\dagger}c_{i\downarrow})$$

$$= \mathbf{I}_{\uparrow} + \mathbf{I}_{\downarrow}. \tag{5}$$

Here, $\mathbf{I}_{\uparrow}$ and $\mathbf{I}_{\downarrow}$ represent the current operators for the up-spin and down-spin electrons, respectively.

To calculate the state current carried by an eigenstate $|\psi_n>$ ($n$ is the state index), we use the operation $<\psi_n|\mathbf{I}_{\mathrm{c}}|\psi_n>$. This eigenstate $|\psi_n>$ can be represented as a linear combination of Wannier states, given by:

$$|\psi_n> = \sum_m (a_{m\uparrow}^n |m\uparrow> + a_{m\downarrow}^n |m\downarrow>) \tag{6}$$

where $a_{m\sigma}^n{}'s$ are the coefficients. The state current for the state $|\psi_n>$ becomes

$$I_c^n = \frac{jet}{N\hbar}\sum_m [e^{-j\theta}\,(a_{m\uparrow}^n)^* a_{m+1\uparrow}^n - h.c.] + \frac{jet}{N\hbar}\sum_m [e^{-j\theta}\,(a_{m\downarrow}^n)^* a_{m+1\downarrow}^n - h.c.]$$

$$= I_{\uparrow}^n + I_{\downarrow}^n \tag{7}$$

where $I_{\uparrow}^n$ and $I_{\downarrow}^n$ are the current components associated with up and down spin electrons, respectively.

To find the net persistent CC in the magnetic ring at absolute zero temperature for a given chemical potential $\mu$, we sum over the lowest energy levels that contribute. Thus, we have

$$\mathbf{I}_{\mathrm{c}} = \sum_n I_c^n. \tag{8}$$

For any finite temperature, the net current expression becomes

$$\mathbf{I}_c = \sum_n I_c^n f(E_n) \tag{9}$$

where $f(E_n)$ is the Fermi-Dirac (FD) distribution function. For non-zero temperatures, we need to take the contributions from all the available states with proper weight factors defined by the FD function.

### 2.2.1 Calculation of spin current

Similar to the charge current operator, the spin current operator is defined as [20, 21]

$$\mathbf{I}_s = \frac{\vec{S}\,\dot{\mathbf{X}}+\dot{\mathbf{X}}\,\vec{S}}{2aN}. \tag{10}$$

In our work, we consider only the z-component, and hence the spin current operator simplifies to

$$I_s^z = \frac{\hbar(\sigma_z\,\dot{\mathbf{X}}+\dot{\mathbf{X}}\,\sigma_z)}{4aN}. \tag{11}$$

Using a similar approach as applied to the CC, the SC for the eigenstate $|\psi_n>$ is obtained as

$$I_s^{z,n} = \frac{\hbar}{2e}(I_\uparrow^n - I_\downarrow^n). \tag{12}$$

For a fixed chemical potential and absolute zero temperature, the total spin current is given by

$$I_s^{z,n} = \sum_n I_s^{z,n}. \tag{13}$$

At any finite temperature, the expression takes the form of

$$I_s^{z,n} = \sum_n I_s^{z,n} f(E_n). \tag{14}$$

## 3 Numerical Results and Discussion

Here, we discuss the essential results of circular charge and spin currents for different input conditions, with a particular focus on their temperature dependence. Since, the charge and spin currents are obtained by adding and subtracting the up and down spin current components, respectively, here in our discussion, we do not present the results of individual current components. Before discussing the results, let us first mention the TB parameter values that are common throughout the discussion. The other parameters that are not constant are mentioned in the appropriate parts. The magnetic moments are assumed to have the same strength, denoted as $h_n = h = 1$. The site energies $\epsilon_{i\uparrow}$ and $\epsilon_{i\downarrow}$ within the ring and $\epsilon_i$ on the chain are set to zero, ensuring a disorder-free system. The NNH strengths $t$ and $\lambda$ are fixed at 1. All the energies are measured in units of eV. The spin current is scaled by the factor $\hbar/2$e. Unless mentioned, the results are computed for zero temperature.

To inspect the specific role of the non-magnetic wire, we start with a setup where the ring is not coupled to the chain, i.e., $\lambda = 0$. For such a situation, the characteristic features of charge and spin currents are shown in Fig. 2, setting $N = 10$ and $\mu = -0.5$. The charge current exhibits a finite variation with the magnetic flux, providing a periodicity of one flux quantum, consistent with previous studies. However, a striking observation is that the spin current remains exactly zero across the entire flux window. The reason behind this vanishing spin current is as follows. A spin-selective phenomenon is observed when a finite mismatch exists between up and down spin energy levels. For $\lambda = 0$, the Hamiltonian of the magnetic ring, decoupled from the chain, can be written as a sum of up and down spin sub-Hamiltonians. Due to the antiparallel arrangements of neighboring magnetic moments these two sub-Hamiltonians become exactly symmetric to each other that can be visualized simply by writing the Hamiltonians matrices associated with up and down spin electrons. Thus, both the up and down spin electronic energy levels are identical, resulting in a vanishing persistent spin current.

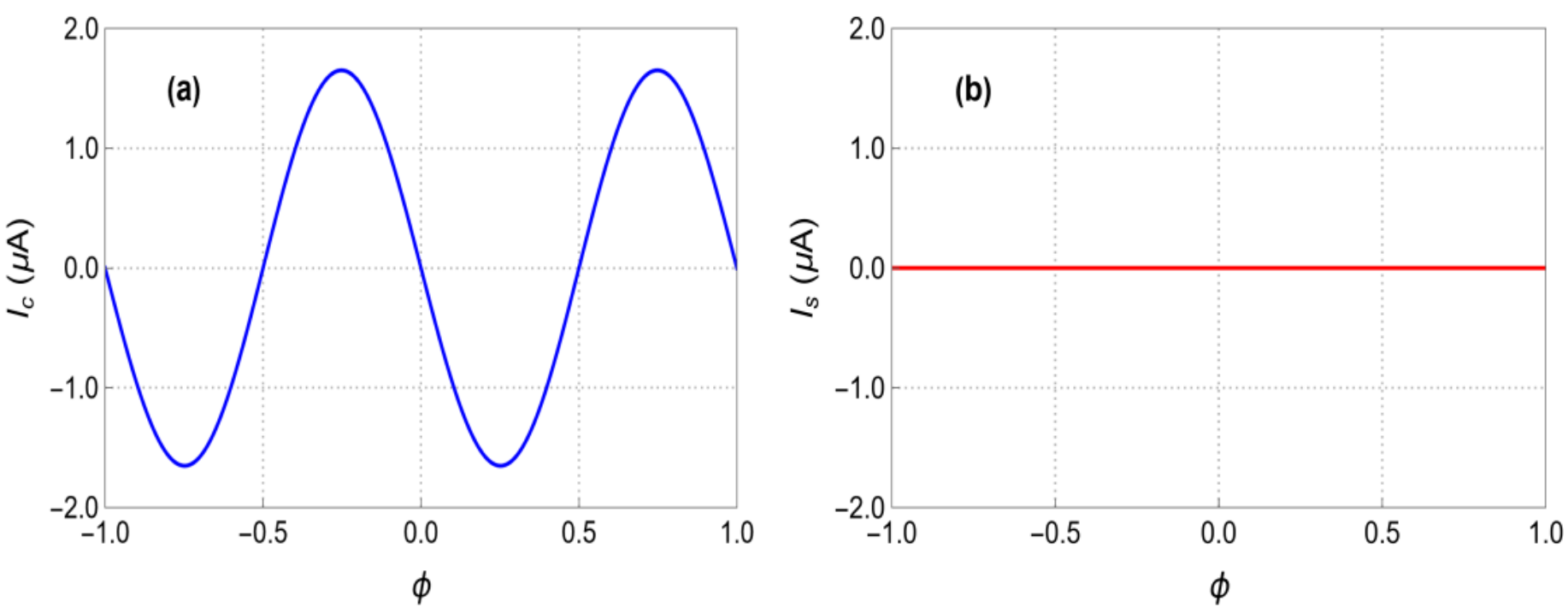

**Fig. 2.** Variation of (a) persistent charge current and (b) persistent spin current in the quantum ring as a function of magnetic flux for $\lambda = 0$. Here we set $N = 10$ and $\mu = -0.5$.

The situation becomes interesting once we couple the ring with the chain. The results of both the charge and spin currents are illustrated in Fig. 3, for $\lambda = 1$. The other physical parameters remain the same as used in Fig. 2. In the presence $\lambda$, the charge current decreases compared to the coupling free case. This reduction of charge current is directly connected to the scattering of electrons at the coupled site of the ring. Because of this coupling, the symmetry between the up and down spin sub-Hamiltonians of the magnetic ring is broken. As a result, a mismatch arises between the two spin-dependent energy channels, leading to a finite spin circular current. The effect of the entire NM chain can be incorporated through renormalization into site α, which then behaves as a disordered site. This disorder effectively breaks the symmetry between the two spin sub-spaces. Although the chain does not carry any net current, it plays a crucial role in generating a finite spin current in the magnetic ring, which itself has zero net magnetization. This is the central focus of our present study. Inspecting the results shown in Fig. 3, it is observed that there is a complete phase reversal between the charge and spin currents. This is solely related to the nature of individual spin current components, as the net charge and spin currents are obtained by adding and subtracting these current components.

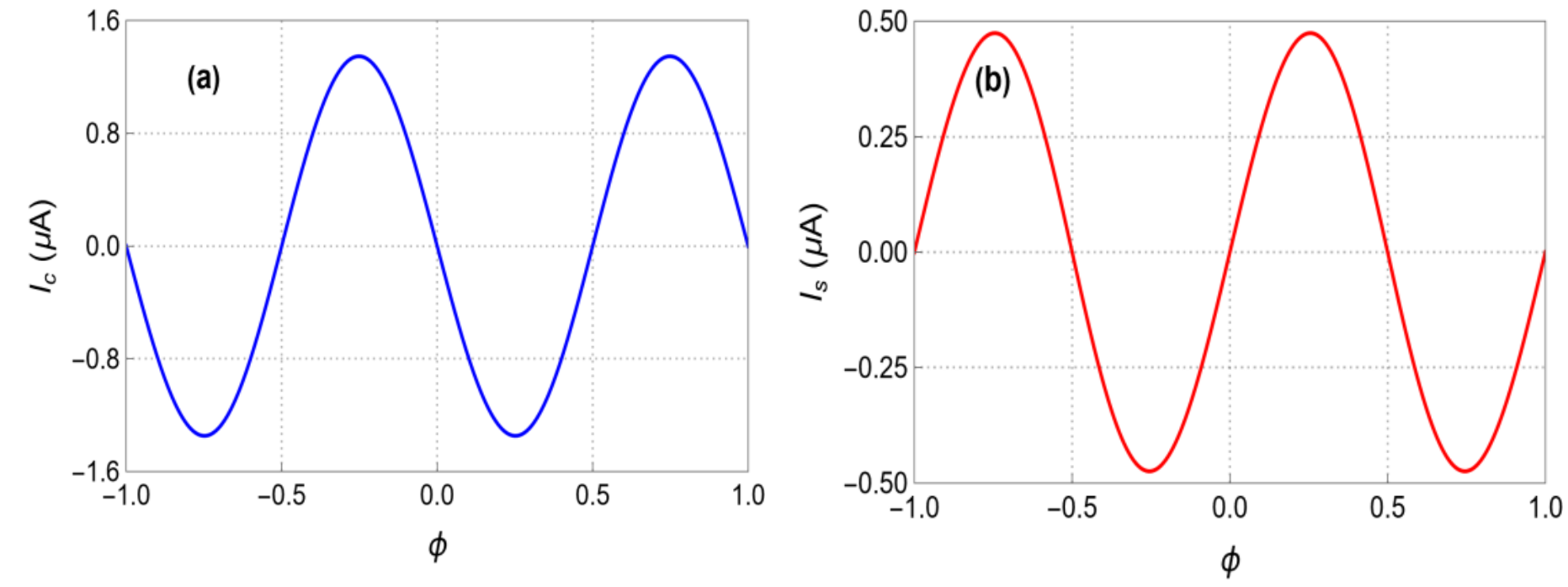

**Fig. 3.** Variation of (a) persistent charge current and (b) persistent spin current in the quantum ring as a function of magnetic flux in presence of finite coupling ($\lambda = 1$) between the ring and the chain. Here we set $N = 10$, $M = 40$, $\alpha = 7$, and $\mu = -0.5$.

It is already established that $\lambda$ has an important role, specifically in the generation of persistent spin current. To clearly demonstrate the dependence of λ on both charge and spin currents, we present in Fig. 4 the variations of these currents over a wide range of λ. Since we are interested only in their magnitudes, the absolute values of the charge and spin currents are shown. The results are quite fascinating. From the variation of the charge current, we observe that it consistently decreases with increasing λ. Enhancing the coupling between the ring and the chain effectively introduces greater disorder in the renormalized site energy α within the ring. This increased disorder leads to stronger scattering, thereby reducing the charge current. The behavior of the persistent spin current with λ, on the other hand, is more intriguing. Initially, the spin current increases with λ, reaches a maximum, and then decreases as the ring-chain coupling strength continues to grow. This non-monotonic behavior can be attributed to the interplay between symmetry breaking and the effective impurity introduced by the coupling. Symmetry breaking is essential for generating spin current, stronger coupling enhances this symmetry breaking, thereby increasing the spin current. However, beyond a critical λ, the enhanced scattering outweighs the benefits of symmetry breaking, resulting in a reduction of the spin current. For sufficiently large coupling, the spin current becomes negligibly small. These results, as illustrated in Fig. 4, clearly indicate that the ring-wire coupling plays a crucial role in tuning both charge and spin currents.

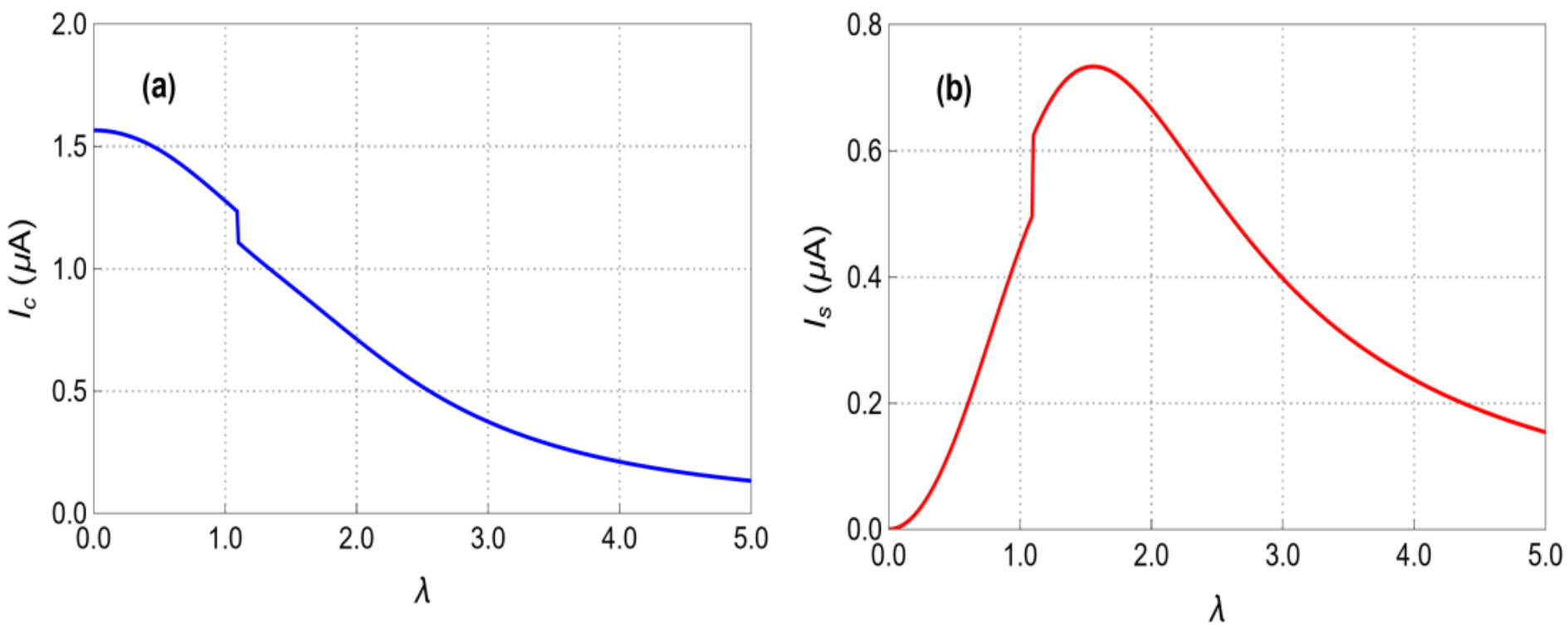

**Fig. 4.** Dependence of (a) persistent charge current and (b) persistent spin current on $\lambda$ in the quantum ring when $N = 10$, $M = 40$, $\mu = -0.5$ and $\phi = 0.2$.

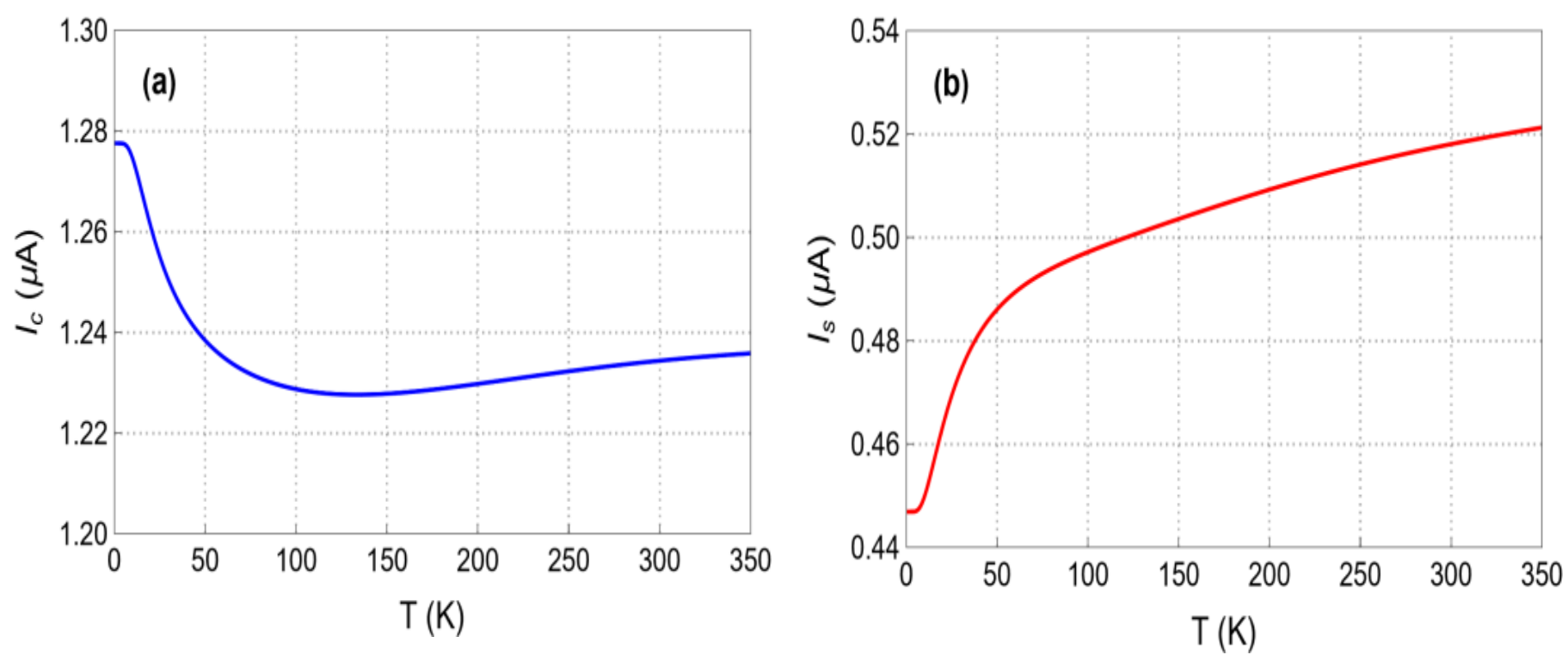

**Fig. 5.** Dependence of (a) persistent charge current and (b) persistent spin current on temperature in the quantum ring when $N = 10$, $M = 40$, $\lambda = 1$, $\mu = -0.5$ and $\phi = 0.2$. Here absolute values of the currents are shown as we are interested in the current magnitudes.

The results discussed so far have been obtained at absolute zero temperature. To account for more realistic conditions, we now examine the effects of finite temperature. In Fig. 5, we present the temperature dependence of both charge and spin currents over a wide temperature range. The observed behaviors are quite unconventional. For the charge current, the magnitude initially decreases within a moderate temperature range but gradually increases beyond that. In contrast, the circular spin current exhibits a monotonically increasing trend throughout the chosen temperature window. These characteristics are closely related to the contributing energy eigenstates, their slopes, and the chosen electrochemical potential ($\mu$). The charge current is derived from the sum of the individual spin current components, while the spin current corresponds to their difference. At low temperatures, the dominant contributions arise from energy levels near $\mu$. As the temperature increases, additional energy levels begin to contribute, enhancing the likelihood of mutual cancellation among channels and thereby reducing the net current. At high temperatures, contributions from all available channels become significant. In this regime, due to the specific choice of μ, one spin component contributes less than the other, resulting in an enhanced charge current. In the case of spin current, the continuous increase with temperature stems from the opposite signs of the two spin current components. The difference between the two current components leads to an overall increase in spin current.

From the results discussed here it is found that charge current is comparatively higher than the spin counterpart. This is not the generic signature; one can also expect higher spin current than the charge current. For a specific setup, the nature of these two currents essentially depends on the mismatch between the two spin-dependent energy channels and the choice of the electro-chemical potential.

## 4 Conclusions

This work presents a study of circular current and spin current in a hybrid system consisting of a non-magnetic chain directly coupled to a magnetic AB ring with zero net magnetization. Spin current emerges when the symmetry between the sub-Hamiltonians is broken through coupling with the non-magnetic ring. In this setup, spin current can also be tuned by adjusting the coupling strength $\lambda$. The temperature dependence of the circular current in this system exhibits intricate behavior.

Before an end, we would like to point out that with the advancement of nanofabrication technologies [22-24], the proposed ring-wire hybrid setup can be realized in a well-equipped laboratory. We strongly believe that the phenomena explored in this study can be experimentally verified under controlled conditions.